\documentclass[telecom,perspective,accept,pdftex,moreauthors]{Definitions/mdpi} 
\usepackage{bbold,braket}

\firstpage{1} 
\pubvolume{6}
\issuenum{3}
\articlenumber{50}
\pubyear{2025}
\copyrightyear{2025}

\datereceived{9 June 2025} 
\daterevised{7 July 2025}
\dateaccepted{10 July 2025} 
\datepublished{14 July 2025} 

\hreflink{https://doi.org/10.3390/telecom6030050} 
\Title{Quantum Perspective on Digital Money: Towards a Quantum-Powered Financial System}
\TitleCitation{Quantum Perspective on Digital Money: Towards a Quantum-Powered Financial System}

\Author{Artur Czerwinski \orcidA{}}
\AuthorNames{Artur Czerwinski}
\AuthorCitation{Czerwinski, A.}

\address[1]{Institute of Physics, Faculty of Physics, Astronomy and Informatics, Nicolaus Copernicus University in Torun, ul. Grudziadzka 5, 87-100 Torun, Poland; aczerwin@umk.pl}

\abstract{Quantum money represents an innovative approach to currency by encoding economic value within the quantum states of physical systems, utilizing the principles of quantum mechanics to enhance security, integrity, and transferability. This perspective article explores the definition and properties of quantum money. We analyze the process of transferring quantum money via quantum teleportation, using terrestrial and satellite-based quantum networks. Furthermore, we consider the impact of quantum money on the modern banking system, particularly in money creation. Finally, we conduct an analysis to assess the strengths and weaknesses of quantum money, as well as opportunities and threats associated with this emerging concept.}


\begin{document}

\section{Introduction}

The concept of money has evolved greatly throughout human history, adapting to the social, technological, and economic contexts of each era. In its earliest form, money was tangible, typically represented by items with intrinsic value, such as gold and silver \mbox{coins \cite{Howgego1995,Holt2021}}. These coins were valued not only as currency but also for their inherent worth as precious metals. Over time, societies began to recognize that the utility of money lay not in its physical substance but in its capacity to represent value. This led to the development of symbolic forms of money, most notably paper bills, which held value not through intrinsic worth but through social and governmental endorsement. This type of money,  commonly known as ``fiat money", carries no economic value but serves solely as a medium of exchange \cite{Wallace2010}. Paper currency revolutionized trade by making transactions simpler and more practical, laying the foundation for modern economies \cite{Read1926}.

With the rise of the digital age, the nature of money shifted once again. Today, much of what we consider ``money" is intangible, represented by numbers in a bank account rather than physical bills or coins. This form of digital currency is managed through secure databases and global financial networks, with value being transferred electronically across great distances in seconds \cite{Allen2004}. While digital money has made transactions faster and more efficient, it has also brought new challenges, particularly in terms of security, privacy, and counterfeiting. Despite rigorous cryptographic techniques, traditional digital money remains vulnerable to certain forms of fraud and cyber-attacks, fueling a search for more secure solutions in the realm of digital finance \cite{Merton1995}.

This ongoing evolution naturally leads to the integration of cutting-edge technologies, such as quantum-enhanced innovations, into financial systems. Quantum-based technologies promise to address several security challenges inherent in current digital finance structures. Among the most intriguing applications of quantum mechanics is the concept of \emph{quantum money}. Quantum money can be formulated based on the principles of quantum information theory, such as the \emph{no-cloning theorem}, which states that it is impossible to create an exact copy of an arbitrary unknown quantum state \cite{Wootters1982}. This fundamental principle enables the conception of a form of money that cannot be counterfeited, as any attempt to duplicate the quantum state representing the money would alter it irreversibly, leading to the destruction of the quantum superposition \cite{Aaronson2009,Aaronson2012,Aaronson2012b,Farhi2012}. This intrinsic security feature has the potential to set quantum money apart from both traditional and digital forms of currency.

This article provides a theoretical exploration of quantum money, discussing its foundational principles, practical implementation, and the challenges that remain. Specifically, in Section~\ref{definition}, we discuss the definition and properties of quantum money and quantum wallets. This concept is elaborated upon in Section~\ref{wallet}, where we explore how quantum wallets can be used without destroying quantum superposition. Next, in Section~\ref{teleportation}, we examine how quantum money could benefit from quantum teleportation for secure transfer across quantum networks, whether via satellite entanglement distribution or terrestrial fiber-based networks. {Moreover, Section}~\ref{Countermeasures} {outlines potential attack vectors against quantum money and proposes corresponding countermeasures to mitigate these threats.} Section~\ref{bankingsystem} presents potential impact of quantum money on modern banking system, with particular focus on the creation of money by commercial banks. Finally, in Section~\ref{SWOT}, we present a SWOT analysis to assess the potential impact, risks, and future directions for quantum money within a quantum-powered financial system.

\section{Defining Quantum Money and Quantum Wallets}\label{definition}

\subsection{Quantum Money}
Quantum money can be defined as symbolic values encoded in quantum states.  Each amount of money would be represented by a unique quantum state, $\ket{\$}$, giving it a ``quantum fingerprint'' that ensures authenticity and prevents forgery. In some quantum money schemes, such as the one proposed in \emph{Quantum Money from Knots} \cite{Farhi2012}, each quantum state is also associated with a classical serial number that can be publicly authenticated, further supporting verification without enabling duplication.

Qubit (quantum bit), the fundamental unit of quantum information, is a typical example of a quantum state. A qubit is a linear superposition of two orthogonal basis states \( \ket{0} \) and \( \ket{1} \), and is described by the combination \cite{Nielsen2000}:
\begin{equation}\label{justqubit}
    \ket{\psi} = \alpha \ket{0} + \beta \ket{1}
\end{equation}
where \( \alpha \) and \( \beta \) are complex numbers, often called the amplitudes of the states \( \ket{0} \) and \( \ket{1} \), respectively. These amplitudes determine the probability of measuring the qubit in either of the two basis states. For example, the probability of measuring the qubit in state \( \ket{0} \) is \( |\alpha|^2 \), and the probability of measuring the qubit in state \( \ket{1} \) is \( |\beta|^2 \). The definition of a qubit in Equation~(\ref{justqubit}) reflects the principle of quantum superposition.

{\textls[-10]{To better understand superposition intuitively, one can think of the famous Schrödinger’s} cat thought experiment. Imagine a cat in a sealed box that can be simultaneously alive and dead until the box is opened and observed. Similarly, a qubit is not just in a state of 0 or 1 but exists in a combination of both states at the same time. It is only when measured that the qubit ‘collapses’ to either 0 or 1 with certain probabilities determined by the amplitudes $\alpha$ and $\beta$. This unique property enables quantum computers and quantum money to perform tasks and provide security features impossible for classical bits.}

Since quantum states are normalized, we require that:
\begin{equation}
|\alpha|^2 + |\beta|^2 = 1,
\end{equation}
which ensures that the total probability of measuring the qubit in one of the basis states is 1. The specific combinations of \( \alpha \) and \( \beta \) would correspond to the different values that quantum money can represent, and these quantum states would serve as the ``signature'' of the money. For instance, one could assign different quantum states (represented by different values of \( \alpha \) and \( \beta \)) to represent different amounts of currency, ensuring a secure and unique encoding of the money.

One of the key properties that makes quantum money particularly secure is its immunity to counterfeiting, which arises from the no-cloning theorem \cite{Wootters1982}. The no-cloning theorem states that it is impossible to make an exact copy of an arbitrary quantum state. In mathematical terms, suppose we have a quantum state $\ket{\psi}$ as in Equation~(\ref{justqubit}), {and an auxiliary system prepared in an initial state $\ket{e}$. The no-cloning theorem asserts that there does not exist a unitary operation $U$ that can clone the state $\ket{\psi}$ in such a way that:
\begin{equation}
U \left( \ket{\psi} \otimes \ket{e} \right) \longrightarrow \ket{\psi} \otimes \ket{\psi}.
\end{equation}

The no-cloning theorem means that a quantum state \( \ket{\psi} \) cannot be perfectly duplicated. If we try to clone it, the result will either be an imperfect copy or a detectable error in the process. This is somewhat analogous to trying to perfectly photocopy a delicate sculpture made of sand—any attempt to reproduce it by touching or measuring it disturbs its shape. Thus, quantum money encoded in unique quantum states cannot be replicated without detection, ensuring its authenticity and preventing forgery. Any attempt to copy the quantum money would either fail or be detectable, making it a highly secure form of currency.

\subsection{Creation of Quantum Money}

To create quantum money, various physical systems can be used for encoding specific quantum states \cite{Messiah1966,Sakurai2020,Paglione2021}. For instance, atoms or ions could serve as the ``carriers'' of quantum money, with their energy levels encoding unique patterns that represent different values. Imagine a single atom of rubidium or calcium, where its specific quantum properties, such as spin or energy states, can be carefully adjusted in a lab to create an identifiable and secure ``signature'' that represents a set amount of money. This atomic state could then be stored and verified using sophisticated quantum devices.

Another possibility involves using photons, the fundamental particles of light, which can be encoded with specific polarization or time-bin states \cite{Scully1997}. These photons could represent quantum money in a secure way because their quantum state is highly sensitive to any attempt at measurement, immediately flagging any interference. For instance, the horizontal or vertical polarization of a photon might encode different monetary values, and these photons could be stored in special quantum memories or transmitted over secure quantum communication channels.

Larger structures like molecules could also encode quantum money, with each molecule’s specific quantum configuration acting as a unique and unforgeable marker. Certain molecules could be designed to have stable quantum states within their electron configurations, creating a ``molecular signature'' that is resistant to tampering. These molecules would be stored in quantum-stable environments to protect the encoded value over time.

In essence, quantum money could be implemented by highly sensitive and complex quantum states---whether in atoms, photons, or molecules---as secure carriers of value, making it possible to create a currency that is nearly impossible to replicate \mbox{without detection.}

\subsection{Protecting Quantum Money from Decoherence}

Quantum states are inherently fragile and can easily lose their distinctive properties due to interactions with their environment---a process known as decoherence \cite{Nielsen2000,Schlosshauer2007}. Decoherence poses a significant challenge for quantum money because it can cause the quantum states encoding the currency to degrade, compromising both their authenticity and security. Even minimal interference from surrounding particles or fields can disturb these states, leading to the gradual loss of the ``quantum fingerprint'' that makes each unit of quantum money unique and secure.

One approach to combating decoherence is to encode quantum money in decoherence-free subspaces (DFS). DFS are specific subspaces within a quantum system where quantum states remain protected from certain types of environmental noise, essentially isolating the information from disruptive influences \cite{Lidar1998}. These subspaces are typically defined in systems that exhibit symmetries, where the environmental noise affects only certain parts of the quantum system and leaves the states in the DFS unaffected.

{One can think of DFS as “quiet zones” or “safe rooms” within a noisy environment, where the quantum information is shielded from the disruptive effects around it, allowing the quantum money’s delicate states to remain stable and reliable.}

Mathematically, DFS can be understood in terms of the interaction between the system and the environment. For instance, suppose a system of two qubits \( \ket{\psi} = \alpha \ket{00} + \beta \ket{11} \) is subject to some form of noise. If the noise affects the qubits symmetrically, then the states \( \ket{00} \) and \( \ket{11} \) can remain invariant under the action of this noise. The key property of DFS is that they encode quantum information in such a way that noise acting on one qubit does not affect the encoded information, which is preserved across the system. 

In general, the noise model acting on a quantum system is described by a superoperator \( \mathcal{E} \). For a state \( \rho \), the action of noise is:

\begin{equation} 
\rho' = \mathcal{E}(\rho)
\end{equation}

For DFS to be effective, the evolution of quantum states within the subspace \mbox{must satisfy}  

\begin{equation}  
\mathcal{E}(\ket{\psi}) = \ket{\psi}  
\end{equation}  
for all states \( \ket{\psi} \) in the DFS, ensuring that these states remain unaffected by the noise.  

For example, a valid DFS under collective bit-flip noise, where both qubits are flipped simultaneously, consists of the Bell states:  

\begin{equation}\label{dfsexample}
\ket{\Psi^+} = \frac{1}{\sqrt{2}} (\ket{01} + \ket{10}) \hspace{0.5cm}\text{and}\hspace{0.5cm} \quad \ket{\Psi^-} = \frac{1}{\sqrt{2}} (\ket{01} - \ket{10}).  
\end{equation}

These states are eigenstates of the collective bit-flip operator \( X^{\otimes 2} \), meaning they remain unchanged up to a global phase under simultaneous flips:

\begin{equation}  
X^{\otimes 2} \ket{\Psi^+} = \ket{\Psi^+}, \quad X^{\otimes 2} \ket{\Psi^-} = \ket{\Psi^-}.  
\end{equation}  

Since the states Equation~(\ref{dfsexample}) are unaffected by collective bit-flip errors, they can serve as an effective DFS for encoding quantum money. By encoding quantum money in DFS, the coherence of quantum states can be preserved for longer periods, making them more stable during storage and transfer.

This increased stability is essential for the practical use of quantum money, ensuring that it maintains its secure quantum state across various processes. Furthermore, DFS can be designed to protect against specific types of decoherence, such as phase noise or amplitude damping, by exploiting symmetries in the noise model. This property allows for the robust encoding and verification of quantum money, making it a promising approach for secure quantum financial systems.

\subsection{Quantum Wallets}

A \emph{quantum wallet} would serve as a secure storage device for quantum money, relying on the principles of quantum memory to maintain the coherence and integrity of quantum states representing currency \cite{Lvovsky2009}. Quantum wallets will be designed not only to hold quantum states but also to ensure their long-term stability against decoherence and interference. To achieve this goal it is essential to implement quantum memory technology, as it preserves quantum information by minimizing interactions with the environment.

In quantum wallets, stability is achieved by storing quantum states in DFS, where specific types of noise do not impact the state. By hosting quantum states in DFS, quantum wallets ensure the durability of quantum money over time. Quantum memory in this context provides the structural support for quantum currency, protecting against rapid decoherence that would otherwise render the stored quantum states unusable \cite{Zhao2009}. For quantum money to be practical, it must retain coherence long enough for standard, everyday transactions, making the robustness of DFS-based quantum memory essential.

{The feasibility of quantum wallets depends on the balance between coherence timescales and transaction speeds. In typical systems, coherence times} (\( \tau_c \)) {of quantum memories can range from microseconds in superconducting qubits to seconds or longer in atomic systems or rare-earth doped crystals. On the other hand, transaction times} (\( \tau_t \))—{which include transmission, verification, and user interaction—must remain well within} \( \tau_c \) {to prevent decoherence. For example, a transaction involving quantum teleportation over satellite links may require several hundred milliseconds, which is feasible only if the coherence time of the wallet exceeds this window. Therefore, practical implementation requires quantum memories with} \( \tau_c \gg \tau_t \), {and ongoing advances in low-noise materials and error mitigation techniques are essential to maintain this condition.}

The use of DFS in quantum wallets will allow quantum money to remain stable and secure over extended periods, much like a traditional currency stored in a vault. The DFS-based quantum memory mitigates noise effects such as collective dephasing, which commonly lead to information loss. This approach is critical for creating a currency system that users can reliably store and access over time, addressing a primary challenge in practical quantum currency applications.

A quantum wallet’s reliance on quantum memory further enhances security, as the no-cloning theorem inherently prevents unauthorized duplication of quantum states. Additionally, any unauthorized attempt to measure or interact with the quantum state would collapse the stored information, rendering it unusable and thwarting counterfeiting attempts. In this way, the quantum wallet provides a tamper-proof solution where only authorized users can access or transfer quantum money without disturbing the state. Therefore, by acting as a stable vault and a security

\subsection{Classical Money vs. Quantum Money}
Traditional forms of money, whether physical or digital, rely on cryptographic or physical measures to prevent duplication and forgery. Quantum money, by contrast, uses intrinsic quantum properties to secure itself. The no-cloning theorem provides a level of security unattainable by classical systems, making quantum money potentially the most secure form of money. Furthermore, quantum money could be stored and transferred within secure quantum wallets, offering a fundamentally new structure for \mbox{monetary systems.}

Furthermore, unlike digital money, which represents only a claim to funds stored in a bank account, quantum money allows for direct possession of value in the form of unique quantum states stored within a quantum wallet. This setup gives users the comfort of actually holding their money, without relying on a bank for storage. With quantum money, ownership is tied directly to the individual’s control over these quantum states, adding a new dimension of personal security and privacy while preserving the advantages of \mbox{digital currency.}

\section{Verification Methods for Quantum Money in Quantum Wallets}\label{wallet}

{Building on the definitions and foundational concepts of quantum money and quantum wallets introduced in Section} \ref{definition}{, a critical aspect for practical implementation is the ability to verify quantum money securely and reliably. Unlike classical money, verifying quantum money is inherently challenging due to quantum mechanical principles such as the no-cloning theorem and state collapse upon measurement. These constraints require innovative verification methods that preserve the integrity of the quantum states while confirming authenticity. In this section, we explore several theoretical approaches to verification — including blind verification protocols, classical tokens paired with quantum states, non-destructive quantum verification, entanglement-based verification, and zero-knowledge proofs — each designed to comply with quantum mechanics and support secure verification in future quantum financial systems.}

\subsection{Blind Verification Protocols}

Blind verification protocols enable quantum wallets to interact with a trusted quantum server or bank to authenticate the amount of quantum money without directly measuring or collapsing the quantum states stored within \cite{Morimae2013,Morimae2014}. In this approach, the quantum wallet can perform a sequence of quantum operations as specified by the server. The results of these operations are then returned to the server, which interprets them to confirm the amount without requiring direct access to the quantum states themselves. By keeping the verification process blind, these protocols maintain the coherence of the states within the wallet and offer a secure verification method.

\subsection{Classical Tokens Paired with Quantum States}

Another approach is to use classical tokens that correspond to each quantum state in the wallet. Each token effectively represents one unit of quantum currency, serving as a count of the stored quantum states. These tokens can be securely stored alongside the quantum money and updated with each transaction, allowing the owner or verifying institution to track the amount of money without accessing the actual quantum states. This classical-quantum pairing enables a non-invasive verification mechanism that avoids direct measurement while preserving the quantum properties of the wallet.

\subsection{Non-Destructive Quantum Verification}

Non-destructive or weak measurement techniques may allow partial verification of quantum money without entirely collapsing the states. In quantum mechanics, measurements typically cause the collapse of a quantum state, reducing the superposition to one of its basis states. Weak measurements, however, are designed to extract limited information about a quantum state while minimizing this collapse. This is achieved by coupling the system to an ancillary meter system, allowing only partial information to be gathered about the system’s state.

Mathematically, weak measurements are described within the framework of weak values, as introduced by Aharonov, Albert, and Vaidman \cite{Aharonov1988}. Unlike projective measurements, weak measurements involve a weak interaction between the system and the measurement apparatus, such that the quantum state is only minimally disturbed \cite{Steinberg2013}.

A weak measurement of an observable \( \hat{O} \) for a quantum state \( \ket{\psi_i} \), followed by post-selection onto a final state \( \ket{\psi_f} \), yields the weak value \cite{Svensson2013}:

\begin{equation}
    O_w = \frac{\bra{\psi_f} \hat{O} \ket{\psi_i}}{\braket{\psi_f|\psi_i}}.
\end{equation}

In a typical weak measurement scenario, if the system is initially in a superposition Equation~(\ref{justqubit}), the weak measurement provides information about both \( \alpha \) and \( \beta \) without fully collapsing the state. The expected shift in the measurement pointer is related to the weak value rather than the standard expectation value.

Weak measurements provide limited information about the system without fully collapsing its superposition, making them useful for quantum verification tasks like checking the presence of quantum money without fully revealing its state. While this approach does not yield exact information on the monetary amount, it can confirm the presence of quantum states with minimal disturbance, offering partial verification that avoids full state collapse. This feature of weak measurement is particularly advantageous in the context of quantum money, where it is crucial to verify authenticity while preserving the integrity of the encoded quantum state.

Building on this concept, methods from quantum computing, such as non-destructive discrimination of Bell states, could also play a critical role in verifying quantum money. Bell states are a class of maximally entangled quantum states that are fundamental to many quantum protocols, including quantum teleportation and quantum key distribution. Techniques used to distinguish Bell states without collapsing their entanglement could be adapted to verify quantum money non-destructively. For instance, if the quantum states representing the money include entangled components or similar quantum properties, protocols for Bell state discrimination could help determine the integrity and amount of quantum money in a wallet.

One notable approach is the gate-based protocol for non-destructive Bell state discrimination proposed by Gupta \cite{Gupta2005,Gupta2007}. This protocol employs quantum gates, such as Hadamard and CNOT gates, to manipulate and distinguish quantum states without destroying the underlying quantum information. By mapping quantum states onto distinguishable configurations, these operations allow for reliable identification without collapsing the superposition or entanglement.

Applying such techniques to quantum money, verification protocols could confirm not only the presence of valid quantum states but also their monetary value, all while preserving the coherence of the quantum wallet. For example, non-destructive discrimination methods could ensure that quantum money remains viable for further transactions or storage, reducing risks associated with state collapse during verification. These capabilities would provide users with an innovative means of validating their quantum currency securely and efficiently, much like checking the balance of a traditional wallet but without risking the integrity of the quantum states.

Moreover, as quantum money is expected to operate within space-based or fiber-based quantum networks, preserving state fidelity during verification is essential. Noise, decoherence, and other environmental factors can degrade the quantum states representing money. Thus, adopting robust non-destructive verification protocols ensures that quantum money remains functional and trustworthy across various storage and transfer scenarios, whether in terrestrial or space-based applications.

\subsection{Entanglement-Based Verification}

In entanglement-based verification, the quantum states within the wallet are entangled with ancillary particles held by a trusted institution, such as a bank. This setup allows the bank to verify the presence and amount of quantum money without directly accessing the quantum states stored in the wallet. The entanglement ensures that the quantum properties of the wallet remain intact while enabling remote verification.

A maximally entangled state between the quantum wallet and the bank's ancillary system can be described using a maximally-entangled Bell state \cite{Nielsen2000,Horodecki2009}:
\begin{equation}\label{Bellstate}
   \ket{\Phi^+} = \frac{1}{\sqrt{2}} (\ket{00} + \ket{11}). 
\end{equation}

In general, an entangled quantum money state shared between the wallet and the bank can be represented as:
\begin{equation}
   \ket{\Psi} = \sum_{i} c_i \ket{i}_W \otimes \ket{i}_B, 
\end{equation}
where \( \ket{i}_W \) are the basis states of the quantum wallet, \( \ket{i}_B \) are the corresponding states held by the bank, and \( c_i \) are complex coefficients satisfying the normalization condition:
\begin{equation}
   \sum_i |c_i|^2 = 1. 
\end{equation}

The verification process relies on quantum correlations between these entangled particles. When the bank measures its part of the entangled system using an appropriate observable \( \hat{O} \), the state of the quantum wallet is projected accordingly. For example, if the bank measures in the computational basis and obtains outcome \( k \), then the state remaining in the wallet collapses to \( \ket{k}_W \). This ensures that any attempt to tamper with the quantum money (e.g., through cloning or unauthorized measurements) will disrupt the entanglement and be detected.

To further enhance security, the verification process can utilize Bell inequalities \cite{Bell1964,Clauser1969} or quantum state tomography to ensure that the quantum correlations remain intact.

By relying on quantum entanglement, this method enables a verification protocol that ensures authenticity while preventing unauthorized duplication or tampering, making it a powerful tool for quantum money authentication.

\subsection{Zero-Knowledge Proofs for Quantum States}

Quantum zero-knowledge proofs (QZKP) provide a sophisticated approach to verification. With QZKPs, the quantum wallet can demonstrate possession of a specific amount of quantum money to a verifier without revealing the actual quantum states. In a zero-knowledge proof, the wallet engages in a protocol with the verifier, allowing them to verify the currency amount while the individual states remain secret and intact. Quantum zero-knowledge protocols have roots in classical cryptography and, when adapted for quantum systems, offer a robust verification method that maintains state confidentiality and coherence.

\section{Integration of Quantum Money with Quantum Networks}\label{teleportation}

Integrating quantum money with emerging quantum networks—whether terrestrial or space-based—raises both technical challenges and promising opportunities, particularly in the domains of security, efficiency, and decentralization of financial transactions. One of the most viable approaches for enabling the transfer of quantum money is quantum teleportation, a protocol that combines quantum entanglement with classical communication to transmit quantum states without physical transport of the underlying particles \cite{Bennett1993}. In this section, we examine the principles of quantum teleportation and outline the infrastructure required to support secure, large-scale quantum money transfers within future quantum communication networks.

\subsection{Quantum Teleportation Mechanism}\label{teleportationmechanism}

Quantum teleportation enables the transfer of a quantum state between two parties, traditionally named Alice (sender) and Bob (receiver), by utilizing a shared entangled pair of qubits and classical communication \cite{Bennett1993,Nielsen2000}, see a conceptual scheme is shown in Figure~\ref{Fteleportation}. Teleportation protocol has been demonstrated experimentally \cite{Boschi1998} and appears a reliable tool for secure money transfer in the quantum-empowered banking system.  When Alice wants to send a quantum state Equation~(\ref{justqubit}) (representing a certain amount of quantum money) to Bob, she begins by performing a Bell-state measurement on the currency state and her part of the entangled pair. This measurement collapses Alice’s part of the entangled state, encoding the information about the currency onto Bob’s entangled qubit through quantum correlations.
\vspace{-12pt}
\begin{figure}[h]
    \includegraphics[width=0.575\columnwidth]{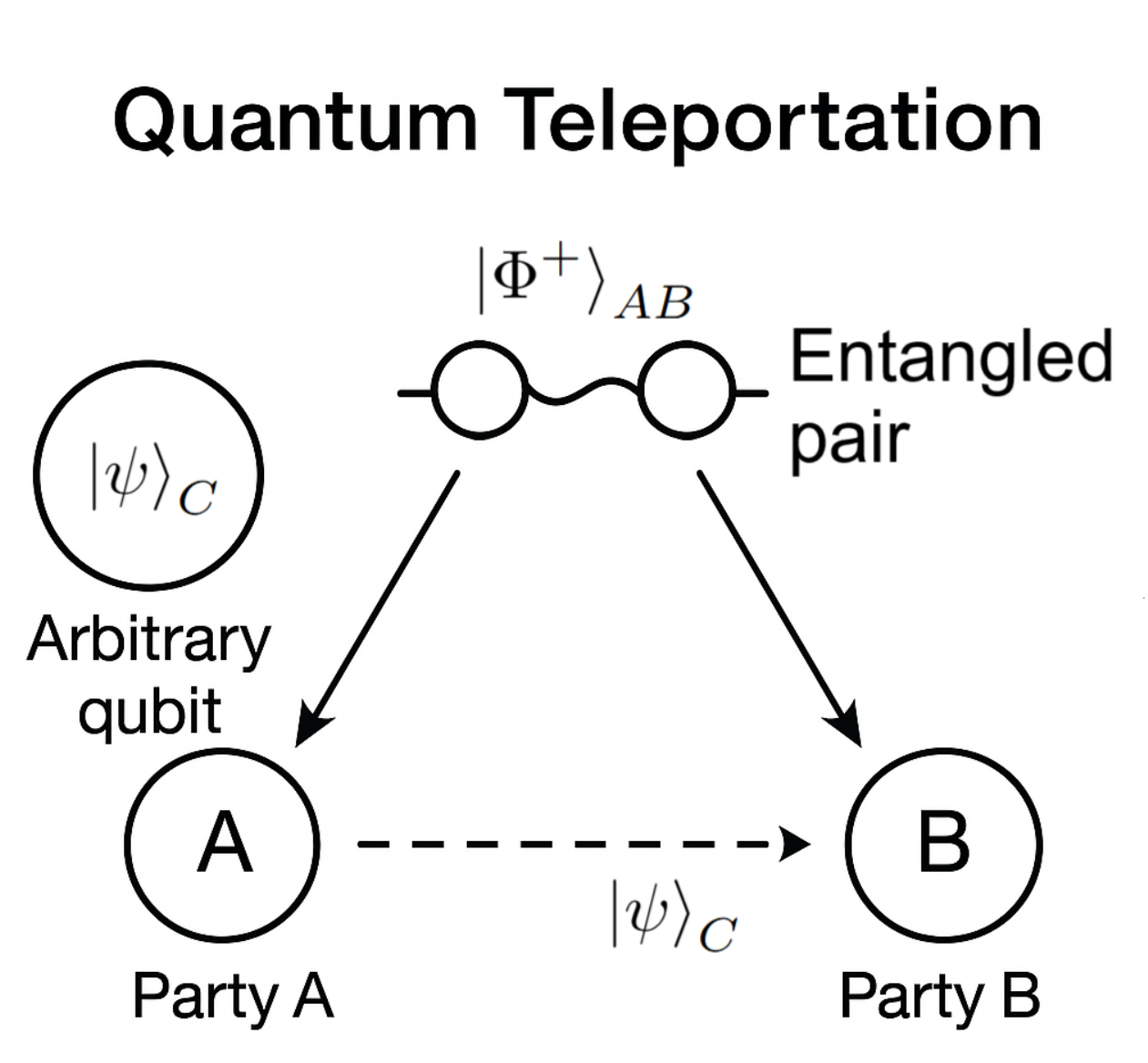}
    \caption{Schematic representation of quantum teleportation utilizing a shared entangled state.}
    \label{Fteleportation}
\end{figure}

Suppose, Alice and Bob share an entangled pair composed of particles $A$ and $B$, initially in one of the Bell states given in Equation~(\ref{Bellstate}). Then, the total system, consisting of the quantum money qubit encoded on particle $C$, as defined in Equation~(\ref{justqubit}), and the entangled pair, is given by:

\begin{equation}
\ket{\psi}_C \otimes \ket{\Phi^+}_{AB} = (\alpha \ket{0}_C + \beta \ket{1}_C) \otimes \frac{1}{\sqrt{2}} \left(\ket{0}_A\otimes\ket{0}_B + \ket{1}_A \otimes \ket{1}_B\right).
\end{equation}

Rewriting this state in the Bell basis for Alice’s two qubits (the money qubit and her half of the entangled pair), we obtain:
\begin{equation}
\begin{aligned}
\ket{\psi}_C \otimes \ket{\Phi^+}_{AB} =& \frac{1}{2} [ \ket{\Phi^+}_{AC} (\alpha \ket{0}_B + \beta \ket{1}_B) + \ket{\Phi^-}_{AC} (\alpha \ket{0}_B - \beta \ket{1}_B)  \\&+ \ket{\Psi^+}_{AC} (\alpha \ket{1}_B + \beta \ket{0}_B) + \ket{\Psi^-}_{CA} (\alpha \ket{1}_B - \beta \ket{0}_B) ].
\end{aligned}
\end{equation}

\textls[+10]{Alice then performs a Bell-state measurement on her two qubits associated with particles A and C. This collapses the state into one of the four Bell states}\linebreak   \( \{ \ket{\Phi^+}_{AC}, \ket{\Phi^-}_{AC}, \ket{\Psi^+}_{AC}, \ket{\Psi^-}_{AC} \} \), each corresponding to a specific transformation of Bob’s qubit.

Alice transmits her measurement outcome (two classical bits) to Bob via a classical communication channel. Based on this outcome, Bob applies one of the following corrective operations to recover \( \ket{\psi}_C \):
\begin{itemize}
    \item If Alice's outcome is \( \ket{\Phi^+}_{AC} \), Bob does nothing.
    \item If Alice's outcome is \( \ket{\Phi^-}_{AC} \), Bob applies \( Z \) (Pauli-Z gate).
    \item If Alice's outcome is \( \ket{\Psi^+}_{AC} \), Bob applies \( X \) (Pauli-X gate).
    \item If Alice's outcome is \( \ket{\Psi^-}_{AC} \), Bob applies \( XZ \) (Pauli-X followed by Pauli-Z gate).
\end{itemize}

Thus, Bob reconstructs the original quantum state $\ket{\psi}_C$, which is a qubit representing the amount of quantum money as in Equation~(\ref{justqubit}).

Importantly, the original quantum state ceases to exist in Alice's system after her measurement, ensuring that the no-cloning theorem is preserved. This makes quantum teleportation an inherently secure method for transferring quantum money without risk \mbox{of duplication.}

\subsection{Benefits of Entanglement-Based Money Transfer}

One key advantage of using quantum teleportation for quantum money transfers is its decentralized nature. Transactions can occur directly between any two parties (such as individuals, banks, or institutions) that share an entangled state. Unlike traditional banking systems, which depend on centralized ledgers or intermediaries, quantum money transfers could theoretically operate on a decentralized quantum network, facilitating peer-to-peer transactions across potentially vast distances. 

Furthermore, since quantum teleportation relies on quantum principles rather than classical cryptography, it is inherently secure against attacks from quantum computers. Quantum computers pose a significant threat to current classical cryptographic systems, but a quantum-enhanced financial system would be resilient to such attacks. This is because its security is derived from quantum mechanics, specifically entanglement and the no-cloning theorem, rather than classical mathematical problems that quantum computers can \mbox{solve efficiently.}

Another benefit of quantum teleportation is the potential for nearly instantaneous transfer of funds. Unlike traditional money transfers, which may take time to process through banks or financial networks, quantum teleportation transfers the quantum state representing money instantly upon receipt of the classical communication. This feature would significantly improve transaction efficiency, with no need to wait for a transfer to clear. As a result, users could experience real-time movement of funds, which could be transformative for applications in global finance.

\subsection{Infrastructure for Quantum Money Transfers}

Implementing a quantum teleportation-based system for money transfers requires a robust quantum infrastructure, utilizing both satellite-based and fiber-optic networks. This infrastructure would distribute entangled pairs to users over distances as needed for transactions. {This approach aligns with the concept of satellite-based Quantum Information Networks (QINs), which connect quantum devices over long distances, significantly enhancing their inherent capabilities in computing, sensing, and security} \cite{Parny2023}.

\subsubsection{Satellite-Based Entanglement Distribution}

For global and long-distance transfers, satellite-based systems could be used to distribute entangled particles over vast distances. This approach is already being explored in quantum key distribution (QKD) and would be suitable for quantum money transfers, connecting users in remote or international locations.

In satellite-based entanglement distribution, pairs of entangled photons are generated and transmitted directly from a satellite—typically orbiting in low-Earth orbit (LEO)—to two optical ground stations. Figure~\ref{Fentanglement} {illustrates this process. Once ground stations A and B share an entangled state, they can perform quantum teleportation of an arbitrary qubit, following the procedure described in Section}~\ref{teleportationmechanism}. {Recent experimental advances have demonstrated the feasibility of satellite-based entanglement distribution between two locations separated by 1203 km on Earth} \cite{Yin2017}.

\begin{figure}[h]
    \includegraphics[width=0.575\columnwidth]{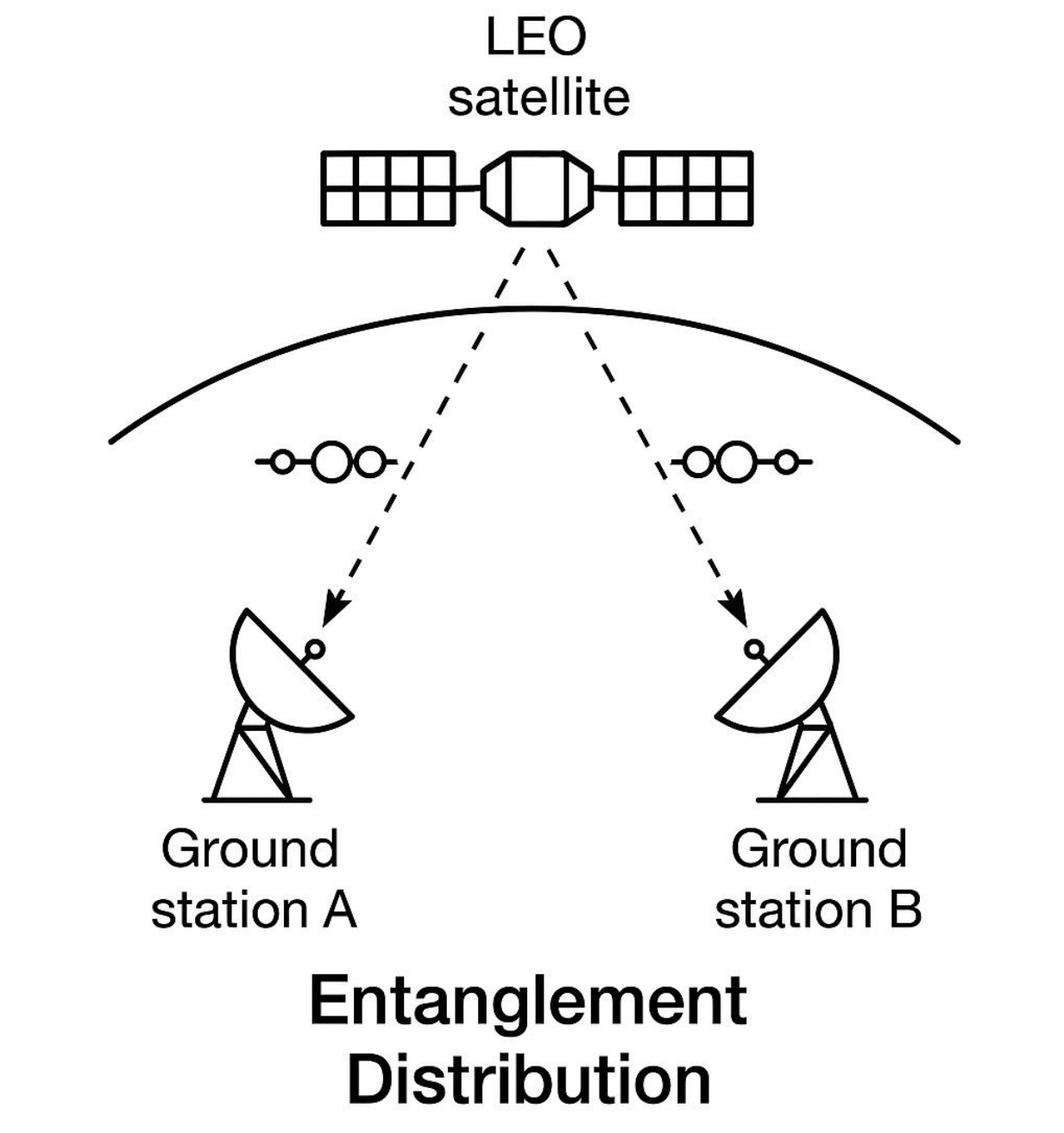}
    \caption{Schematic representation of satellite entanglement distribution to two ground stations.}
    \label{Fentanglement}
\end{figure}
    
\subsubsection{Fiber-Based Networks}

For shorter-range or regional transactions, fiber-optic networks could provide a secure means of distributing entangled states in urban or local areas. Fiber-based systems would enable fast and reliable entanglement distribution, making quantum money transfer feasible within regions served by such infrastructure.

Recent advancements have significantly progressed towards the realization of complex QINs capable of handling multiple use cases, including quantum money operations. For instance, in 2022, entanglement of single atoms was demonstrated over a 33-km telecom fiber link \cite{Leent2022}, while in 2024, entanglement between two nuclear spin memories was achieved through a 40-km low-loss fiber link \cite{Knaut2024}.

\subsection{Large-Scale Quantum Money Operations via Distributed Quantum Computing}

To enable large-scale operations with quantum money, such as interbank transactions, quantum teleportation can be extended through distributed quantum computing. In this scenario, entanglement is distributed between quantum computers located in different banks, allowing quantum money transfers to be performed securely and efficiently.

\textls[-5]{A promising approach involves the use of satellite-based entanglement \mbox{distribution \cite{Yin2017}}}. A satellite can generate and distribute pairs of entangled photons directly between \mbox{two banking} institutions, as illustrated in Figure~\ref{Fentanglement}. Alternatively, this distribution can be facilitated by a third party, known as a quantum repeater, where entanglement between the two institutions is established through entanglement swapping \cite{Azuma2023}. Each institution stores its part of the entangled pair in a quantum memory. When a transfer is requested, quantum teleportation can be performed using the pre-shared entanglement, allowing secure movement of quantum money across distant locations.

Recent advancements have shown that distributed quantum computing can be successfully performed over an optical link \cite{Main2025}. By combining quantum teleportation with distributed quantum computing, financial systems can ensure high security and efficiency in quantum money transactions, paving the way for future quantum-based financial networks.  In this scenario, operations involving quantum money could be performed within the Quantum Internet---a space-based network interconnecting quantum devices to enable secure communication and high-capacity computing. Satellite-based implementations for various use cases of the Quantum Internet have already been proposed; see, for example, Ref.~\cite{Khatri2021}.

\subsection{Quantum Blockchain Integration}

To ensure the integrity and non-repudiation of quantum money transactions, a quantum blockchain can be employed as a decentralized ledger that securely records each transaction within a quantum-powered financial system. Unlike classical blockchains, which often rely on energy-intensive consensus protocols such as proof-of-work, quantum blockchains rely on quantum cryptographic protocols and quantum communication to achieve consensus more efficiently and with enhanced security \cite{Li2019}.

{Quantum blockchains utilize quantum-secure consensus algorithms based on entanglement and quantum communication channels, which are inherently resistant to attacks by quantum computers} \cite{ernandez2020}. {To fully realize their potential, it is necessary not only to replace classical communication with quantum channels but also to develop new consensus protocols suited to quantum environments. In this context, the recently proposed Time-Bin Conference Key Agreement (TB-CKA) protocol offers a promising solution to the consensus problem in distributed systems} \cite{Misiaszek2023}. {This protocol is designed to operate effectively within large-scale QINs.}

{Notably, quantum blockchain consensus mechanisms can incorporate quantum algorithms such as boson sampling or quantum amplitude amplification to significantly lower the computational complexity of transaction validation. Unlike classical mining processes, which are computationally intensive and energy-demanding, quantum protocols can achieve consensus with substantially reduced resource consumption. This advancement directly addresses one of the most critical limitations of classical blockchain systems—namely, their high energy inefficiency—by offering a more sustainable and scalable alternative} \cite{Kearney2023}.

{By replacing classical energy-intensive consensus with quantum-secure, efficient quantum protocols, quantum blockchains offer a potential for sustainable distributed ledger technology that complements and enhances the quantum money ecosystem.}

\section{Security of Quantum Money: Attack Vectors and Countermeasures}\label{Countermeasures}

{The security of quantum money relies fundamentally on the principles of quantum mechanics, such as the no-cloning theorem and quantum entanglement. However, the practical implementations of quantum money systems will inevitably face a range of potential attack vectors. Identifying and mitigating these threats is critical to ensuring the robustness and trustworthiness of quantum financial infrastructure.}

\subsection{Potential Attack Vectors}

\begin{itemize}
\item Physical tampering and hardware attacks: 
 {As with any technology relying on physical devices, quantum wallets and quantum memories may be vulnerable to physical tampering. This includes invasive attacks aimed at extracting stored states or damaging memory coherence.}
  
\item Side-channel attacks: 
 {Imperfections in device design may allow adversaries to infer information about quantum states through measurable side effects, such as timing, power consumption, or emitted radiation.}

\item Spoofing and impersonation: 
 {If a malicious actor is able to intercept classical metadata (e.g., serial numbers or digital certificates), they might attempt to present a fake quantum state that mimics valid money, especially if the verification procedure is not sufficiently robust.}

\item Attacks on quantum communication channels: 
 {During quantum teleportation or entanglement distribution, an adversary may attempt a man-in-the-middle attack, injecting noise, faking entangled pairs, or disrupting transmission fidelity.}

\item Forgery via imperfect verification: 
{If the verification algorithm $M$ for quantum money accepts invalid states with non-negligible probability, adversaries could attempt to produce counterfeit quantum states that pass as legitimate currency.}
\end{itemize}

\subsection{Countermeasures and Defensive Strategies}

\begin{itemize}
\item DFS: 
 {Quantum wallets should encode quantum money in DFS to shield stored states from environmental noise and collective decoherence, which not only extends storage times but also resists certain classes of tampering.}
\item Tamper-evident quantum hardware: 
 {Devices used to store and process quantum money must include tamper detection mechanisms. Unauthorized access or manipulation should result in the collapse of the quantum state, rendering the stored money unusable.}

\item Authentication of classical data: 
 {Quantum money may be associated with classical serial numbers or metadata. These identifiers should be cryptographically authenticated using digital signatures to ensure that only authorized entities (e.g., the central mint) can issue valid quantum tokens.}

\item Redundant verification protocols: {Instead of relying solely on a single measurement outcome, quantum money verification protocols should include probabilistic checks or interactive procedures that increase the difficulty of forging valid states without detection.}

\item Quantum channel authentication: 
 {Communication over quantum networks (used for teleportation or entanglement distribution) must be authenticated to prevent spoofing or hijacking. This includes techniques from QKD to generate secure encryption keys and classical cryptographic authentication for control messages.}

\item \textls[-15]{Device-independent protocols: 
 {Where feasible, device-independent approaches—where} security does not depend on trusting the internal workings of the hardware—can provide added resilience against implementation-level attacks.}
\end{itemize}

{Quantum money offers inherent security advantages due to its non-clonability and measurement sensitivity. Nonetheless, ensuring its practical deployment requires a comprehensive security model that addresses not only quantum information-theoretic guarantees but also physical and cryptographic safeguards. By incorporating these countermeasures, a functional and resilient infrastructure for quantum money can be achieved within emerging quantum networks.}

\section{Impact of Quantum Money on the Banking System}\label{bankingsystem}

The current financial system relies heavily on the ability of commercial banks to create money through debt \cite{McLeay2014,Botos2016}. When a bank issues a loan, it does not lend out pre-existing funds; instead, it creates a new deposit, effectively generating money through accounting entries. This process results in a majority of money existing only as digital records, rather than being backed by physical assets. Consequently, commercial banks exert significant influence over the money supply, often surpassing the central bank’s control.

This section explores the hypothesis:
\begin{quote}
{\it If non-replicable quantum money replaces digital deposits, banks would stop creating money through debt, and central banks would have full control over money supply.}
\end{quote}

{Our approach to quantum money assigns to the central bank the role of the sole trusted institution authorized to create money. This raises the issue commonly known as the trusted institution paradox. On one hand, quantum money aims to decentralize financial transactions (e.g., through quantum teleportation) and prevent unauthorized duplication through the fundamental laws of quantum mechanics. On the other hand, the system inherently depends on a single trusted authority (the central bank) to certify and issue valid quantum currency. Rather than viewing this dependence as a weakness, our approach highlights that quantum money actually reinforces and strengthens the central bank’s critical function in monetary control. The paradox is thus resolved by recognizing that quantum money enhances the central bank’s role as the trusted issuer, while eliminating the ability of commercial banks or other entities to expand the money supply arbitrarily through money creation. This unique balance preserves necessary trust in the national currency but harnesses quantum mechanics to secure and limit money creation strictly to authorized institutions.}

\subsection{The Problem of Symbolic Money Creation}

In the contemporary banking system, money is predominantly a symbolic construct, existing as digital values in bank accounts. Since commercial banks create money when issuing loans, this money supply expansion is not directly tied to real economic value. The mechanism can be described as follows: when a bank grants a loan of amount $M_{loan}$, it simultaneously creates a liability in the form of a deposit of the same amount $M_{loan}$, increasing the total money supply. Mathematically, this can be written as:
\begin{equation}
M_{total} = M_{existing} + M_{loan},
\end{equation}
where $M_{total}$ represents the expanded money supply after loan issuance.

The issue with this process is that it is fundamentally based on trust: money only functions because individuals believe in its value and the banking system's ability to honor deposits \cite{Goldberg2005}. Furthermore, excessive money creation through loans can lead to financial instability and inflation.

\subsection{Quantum Money as a Constraint on Money Creation}

Quantum money introduces a paradigm shift by ensuring that every unit of currency is represented by a specific quantum state. Due to the no-cloning theorem \cite{Wootters1982}, quantum money cannot be duplicated, preventing the arbitrary expansion of the money supply. If banks were required to hold actual quantum money before issuing loans, they would no longer be able to create deposits purely through accounting. This would enforce a fundamental constraint:
\begin{equation}
M_{new} \leq M_{quantum},
\end{equation}
where $M_{quantum}$ represents the total supply of quantum money, directly controlled by the central bank.

This shift would restore full monetary control to the central bank, ensuring that money issuance is explicitly tied to tangible, non-replicable quantum assets. With this approach, only the central bank or other authorized and properly equipped institutions would be able to generate certified quantum states representing legitimate money. These quantum states, recognized as a legal means of payment, could then be transferred to recipients via quantum teleportation.

\subsection{Implications for Financial Stability}

By eliminating arbitrary deposit creation, quantum money could introduce several stabilizing effects:
\begin{itemize}
    \item Elimination of uncontrolled credit expansion:  Since banks could only lend what they actually possess in quantum money, the risk of credit bubbles would be mitigated.
    \item Stabilization of inflation and preservation of savings: 
 By restricting money creation to authorized institutions, quantum money could prevent excessive monetary expansion, reducing inflationary pressures and ensuring that savings retain their value \mbox{over time.}
    \item Direct central bank oversight: 
 The money supply would be entirely controlled by the central bank, eliminating the disproportionate influence of commercial banks.
\end{itemize}

\subsection{Challenges and Future Considerations}

While quantum money could provide a more stable and transparent financial system, its implementation would face challenges, including:
\begin{itemize}
    \item Technological infrastructure: 
 The need for quantum-secured financial networks to store and transfer quantum money.
    \item Resistance from commercial banks: 
 Given their reliance on debt-based money creation, commercial banks may oppose such a transformation.
    \item Economic growth considerations: 
 Some argue that restricting money creation could slow down economic expansion by limiting credit availability.
\end{itemize}

In a future where large-scale quantum networks enable distributed quantum computing, quantum money transactions could be facilitated by entanglement-based transfers between financial institutions. For instance, quantum entanglement could be distributed between two quantum computers located in banks via satellite links, ensuring secure, tamper-proof quantum transactions. Such a system could redefine monetary policy and financial stability on a global scale.

The replacement of traditional digital deposits with quantum money could fundamentally alter the structure of the banking system. By enforcing physical constraints on money supply expansion, quantum money could restore central banks’ control over the economy and prevent financial excesses caused by unchecked credit creation. Future research should focus on the feasibility of large-scale quantum money deployment and its implications for global finance.

\subsection{Comparison with Central Bank Digital Currencies (CBDCs)}\label{cbdccomparison}

{According to the definition, a Central Bank Digital Currency (CBDC) is a digital form of currency issued by a central bank and constitutes a liability of that issuing authority} \cite{Kiff2020,Ozili2023}. {This subsection aims to compare and contrast the key features of quantum money and CBDCs. Although both seek to modernize the financial system and offer more secure digital payment methods, they differ fundamentally in their operational principles, underlying technologies, and potential impacts on monetary policy.}

{CBDCs are digital tokens issued by central banks as legal tender, typically stored in conventional databases or digital wallets. They aim to provide a state-backed alternative to private digital payment systems, offering enhanced payment efficiency, programmability, and traceability. However, CBDCs do not address the core structural issue of symbolic money creation, as they remain classical information objects that can be copied and manipulated within authorized systems. Moreover, in most proposed designs, CBDCs coexist with traditional deposit-based money, leaving the credit-creation function of commercial banks intact} \cite{Bibi2024}.

{By contrast, quantum money introduces a physical limit on money creation, as each unit is encoded in a specific quantum state that cannot be duplicated due to the no-cloning theorem. This fundamentally constrains the creation of new currency units, directly tying money supply to quantum physical processes and removing the symbolic flexibility exploited in conventional and even CBDC frameworks. If implemented at scale, quantum money would act not only as a secure medium of exchange, but also as a structural reform tool to prevent the uncontrolled expansion of credit.}

{In summary, as presented in Table}~\ref{tab:cbdc_vs_quantum}{, while CBDCs represent a digital evolution of existing monetary models, quantum money challenges the foundational assumptions of money itself. Its implementation would not merely digitize central bank currency, but redefine monetary sovereignty and structural stability in a future quantum-enabled~economy.}

\begin{table}[H]
   \small
\caption{Comparison between CBDCs and Quantum Money.}
\label{tab:cbdc_vs_quantum}

\setlength{\cellWidtha}{\textwidth/3-2\tabcolsep-0in}
\setlength{\cellWidthb}{\textwidth/3-2\tabcolsep-0in}
\setlength{\cellWidthc}{\textwidth/3-2\tabcolsep-0in}
\scalebox{1}[1]{\begin{tabularx}{\textwidth}{>{\raggedright\arraybackslash}m{\cellWidtha}>{\raggedright\arraybackslash}m{\cellWidthb}>{\raggedright\arraybackslash}m{\cellWidthc}}
\toprule

\textbf{Feature} & \textbf{CBDC} & \textbf{Quantum Money} \\
\cmidrule{1-3}
Technological foundation  & Classical digital information systems & Quantum states governed by quantum mechanics \\
\cmidrule{1-3}
Copyability
 & Can be duplicated under institutional control & Fundamentally uncopyable due to no-cloning theorem \\
\cmidrule{1-3}
Money creation mechanism
 & Issued by central bank, but often coexists with symbolic money created by banks & Each unit must be physically created and distributed by authorized entities using quantum systems \\
\cmidrule{1-3}
Security
 & Vulnerable to classical cyberattacks; depends on encryption & Intrinsically secure; tampering collapses the state and renders it invalid \\
\cmidrule{1-3}
Integration with quantum infrastructure
 & Not inherently designed for quantum networks & Native use case for satellite and fiber-based QINs (e.g., teleportation, entanglement-based transfers) \\
\cmidrule{1-3}
Impact on commercial banks
 & May allow commercial banks to continue creating digital money through loans & Constrains money creation to physical issuance; limits banks' ability to create money via accounting \\
\bottomrule
\end{tabularx}}

\end{table}

\subsection{Hybrid Quantum-Classical Systems as Interim Solutions}\label{hybridwallets}

While fully quantum money systems remain technologically demanding, hybrid quantum-classical systems can serve as a practical bridge between classical financial infrastructures and emerging quantum technologies. In such systems, quantum protocols would be integrated into conventional banking operations to enhance security and provide a gradual path toward full quantum financial systems.

In the early stages of adoption, quantum teleportation can be employed not to transfer quantum money itself, but rather to transmit quantum information—such as a validation token or verification state—between parties. This quantum state, encoded in a qubit and securely teleported, could be used to authorize or validate classical financial transactions without being susceptible to eavesdropping or duplication. In this configuration, the quantum component supports authentication and verification, while the actual transfer of monetary value remains within the classical domain.

Moreover, QKD protocols could play a crucial role in hybrid systems by enabling the generation of truly secure encryption keys for protecting sensitive financial \mbox{communications \cite{Pirandola2020}}. {With QKD protocols, banks and financial institutions could establish shared cryptographic keys over quantum channels, thereby eliminating vulnerabilities associated with classical key exchange methods. These quantum-generated keys would then be applied in standard encryption protocols (e.g., AES) used in the existing digital banking framework.

Hybrid wallet architectures would therefore allow institutions to incrementally incorporate quantum technologies without overhauling their entire infrastructure. By enabling secure quantum-based verification and communication, these systems would enhance the robustness of financial operations while paving the way for future integration of fully quantum monetary units. Such an approach aligns with realistic technological progress and helps to build trust and expertise before a complete transition to quantum-native finance becomes feasible.}

\section{SWOT Analysis of Quantum Money}\label{SWOT}

SWOT analysis is a strategic planning tool used to identify and assess the key factors influencing the success or viability of a project, product, or concept. Standing for \textbf{S}trengths, \textbf{W}eaknesses, \textbf{O}pportunities, and \textbf{T}hreats, SWOT analysis provides a structured framework that allows for a comprehensive evaluation from both internal and external \mbox{perspectives \cite{Weihrich1982}.} By categorizing these factors, SWOT analysis helps stakeholders recognize potential advantages, mitigate risks, and make informed decisions. Strengths and weaknesses focus on internal aspects, highlighting the inherent benefits and limitations, while opportunities and threats are external, encompassing broader factors like market trends, technological advancements, and competitive pressures.

The primary goal of SWOT analysis is to support effective decision-making and strategic development by creating a balanced view of current circumstances and future possibilities. By understanding strengths, an organization or team can utilize them to enhance their offering; by acknowledging weaknesses, they can identify areas that may need improvement. Opportunities represent favorable external conditions that could support growth or success, whereas threats pose potential challenges or risks. Taken together, SWOT analysis guides organizations in aligning their resources and strategies to achieve competitive advantage, anticipate changes, and optimize outcomes. Table~\ref{swotana} presents the SWOT analysis for the concept of quantum money.

\begin{table}[H]
   \small
   \caption{SWOT analysis of quantum money.}\label{swotana}
\label{swotana}

\setlength{\cellWidtha}{\textwidth/2-2\tabcolsep-0in}
\setlength{\cellWidthb}{\textwidth/2-2\tabcolsep-0in}
\scalebox{1}[1]{\begin{tabularx}{\textwidth}{>{\raggedright\arraybackslash}m{\cellWidtha}>{\raggedright\arraybackslash}m{\cellWidthb}}
\toprule

\textbf{Strengths} & \textbf{Weaknesses} \\ \cmidrule{1-2}
No-cloning theorem provides unmatched security (uncounterfeitability of quantum money) & Sensitive to decoherence and noise (e.g., atmospheric turbulence in free-space transmission) \\ \cmidrule{1-2}
Quantum teleportation prevents duplication during transfer & High costs of developing world-wide quantum infrastructure \\ \cmidrule{1-2}
Decentralized and secure through quantum blockchain & Requires reliable and high-fidelity entanglement distribution \\ \cmidrule{1-2}
Robust against classical and quantum attacks & Challenges of compatibility with existing banking infrastructure \\ \cmidrule{1-2}
Encoded in specific physical quantum systems, making it a tangible form of money with the potential to restrict unrestricted money creation by commercial banks. & Quantum wallets may be costly to produce and could have limited portability compared to traditional digital banking solutions\\
\bottomrule
\end{tabularx}}
\end{table}

\begin{table}[H]\ContinuedFloat
\tablesize{\small}
\caption{\textit{Cont.}}
\setlength{\cellWidtha}{\textwidth/2-2\tabcolsep-0in}
\setlength{\cellWidthb}{\textwidth/2-2\tabcolsep-0in}
\scalebox{1}[1]{\begin{tabularx}{\textwidth}{>{\raggedright\arraybackslash}m{\cellWidtha}>{\raggedright\arraybackslash}m{\cellWidthb}}
\toprule

\textbf{Opportunities} & \textbf{Threats} \\ \cmidrule{1-2}
Potential for global adoption with a robust Quantum Internet (currently being developed by ESA and other agencies) & Technical challenges in maintaining coherence \\ \cmidrule{1-2}
Could redefine financial systems through advanced security & Possible vulnerabilities if quantum error correction fails \\ \cmidrule{1-2}
Opportunity to protect against quantum computer attacks on classical systems & Regulatory and legal obstacles to adoption \\ \cmidrule{1-2}
Positive boost for the economy by making global transactions easier and faster & Potential misuse for financing illegal operations \\ \cmidrule{1-2}
Potential to strengthen the role of the central bank in controlling inflation by regulating the money supply & Possible negative impact on economic growth due to restrictions on money creation.\\

\bottomrule
\end{tabularx}}
\end{table}

\section{Conclusions}

Quantum money offers an exciting vision for the future of digital currency, with security features grounded in quantum mechanics that surpass classical capabilities. By relying on the no-cloning theorem, quantum teleportation, and entanglement distribution, quantum money could transform financial systems, enhancing both security and decentralization. While some aspects of quantum money have already been discussed \cite{Aaronson2009,Aaronson2012,Aaronson2012b,Farhi2012}, the present perspective provides new insights into the practical implementation of quantum wallets, the potential impact of quantum money on the financial system (including a comparison with CBDCs), and its integration within emerging concepts of the Quantum Internet. Moreover, we showcase key features of quantum money by performing a SWOT analysis, which highlights not only its positive aspects but also the challenges associated with this concept.

Realizing the potential of quantum money will require overcoming technical, infrastructural, and regulatory challenges. In addition, one might expect social reluctance to adopt quantum money. Already it can be seen that the widespread use of digital currency prompts some people to defend the right to use cash. However, with ongoing advancements in quantum technology, especially in the Quantum Internet, quantum money may one day become a practical and highly secure form of currency, integrated with currently developing networks for quantum communication and computation.

The use of quantum teleportation for transferring quantum money promises a secure, efficient, and decentralized alternative to conventional banking transactions. With continued advancements in quantum infrastructure and the integration of quantum blockchains, a quantum financial system could redefine global finance, benefiting from the fundamental principles of quantum mechanics for enhanced security and real-time transfers.  Moreover, since quantum money is intrinsically tied to specific physical particles used to encode quantum states, its creation will be restricted to authorized institutions capable of generating certified states. This will inevitably limit the widespread money creation by commercial banks, helping to reduce inflation and better preserve the value of savings.

The concept of quantum money is further enriched by advances in quantum communication and network infrastructure. The advent of the Quantum Internet and quantum-secure networks offers a framework for storing, validating, and transferring quantum money over long distances, potentially reshaping the financial landscape. By integrating quantum money with these emerging infrastructures, it may be possible to create a decentralized, highly secure financial system. Furthermore, combining quantum money with quantum blockchain technology could lead to a new form of ledger that not only records transactions but also ensures that all records are protected against tampering and unauthorized duplication. Since entanglement-based quantum technologies are transitioning from laboratories to real-world applications, we can expect that transferring quantum money will someday be just an additional use case for QINs.
\vspace{6pt}

\funding{This research received no external funding.}

\dataavailability{No new data were created or analyzed in this study. Data sharing is not applicable to this article.} 

\conflictsofinterest{The authors declare no conflicts of interest.}

\begin{adjustwidth}{-\extralength}{0cm}
\reftitle{References}

\PublishersNote{}
\end{adjustwidth}
\end{document}